\documentclass{ws-ijmpa}

\begin{document}

\markboth{M. Silarski}
{STUDY OF THE $K^+K^-$ INTERACTION AT COSY-11}

%
\catchline{}{}{}{}{}
%

\title{STUDY OF THE $K^+K^-$ INTERACTION AT COSY-11}

\author{MICHA{\L} SILARSKI
}

\address{Institute of Physics, Jagiellonian University, PL-30-059 Cracow, Poland\\
Michal.Silarski@lnf.infn.it}

\maketitle


\begin{abstract}
In this article we present studies of the near threshold $pp\to ppK^+K^-$ reaction
in view of the $K^+K^-$ final state interaction. The investigations include analysis of both the
low-energy $K^+K^-$ invariant mass distributions measured by COSY-11 collaboration at excess
energies of Q = 10 MeV and Q = 28 MeV and the near threshold excitation function for the
$pp\rightarrow ppK^{+}K^{-}$ reaction.
As a result of these studies we have estimated the $K^+K^-$ scattering length more precise
compared to the previous analysis based only on the analysis of the differential cross sections.
\end{abstract}
\ccode{PACS numbers: 13.75.Lb, 13.75.Jz, 25.40.Ep, 14.40.Aq}
\section{Introduction}
Studies of the hadronic interaction of the $K^+$ and $K^-$ mesons is of a great importance especially regarding
the unknown nature of the $a_{0}$(980) and $f_{0}$(980) scalar resonances, which besides the standard
interpretation as a quark-antiquark state~\cite{Morgan} were also proposed to be a $K\bar{K}$ molecules~\cite{Lohse,Weinstein}.\\
Due to lack of the kaonic targets we investigate experimentally the strength of the $K^+K^-$ interaction
for example in the near threshold $pp\to ppK^+K^-$ reaction where kaons are produced with low relative momenta~\cite{review}.\\
Measurements of the $pp\to ppK^+K^-$ reaction at the kinematical threshold have been made possible by beams of low emittance and small
momentum spread available at storage ring facilities, in particular at the cooler synchrotron COSY at the research center
in J{\"u}lich, Germany~\cite{cosy}.
Early experiments at COSY conducted by means of the  COSY-11 detector~\cite{brauksiepe} revealed, 
however, that a possible influence
from the $f_{0}$ or $a_{0}$ mesons on the $K^{+}K^{-}$ pair production is too weak to be distinguished from
the direct production of these mesons~\cite{quentmeier}.
However, the combined systematic collection of data below~\cite{quentmeier,wolke,winter}
and above~\cite{anke,disto} the $\phi$ meson threshold reveal a significant signal
in the shape of the excitation function which  may be due to the $K^{-}p$ and perhaps
also to the $K^{+}K^{-}$ interaction.
\section{Study of the dynamics in the low energy $ppK^+K^-$ system}
The measurements of the $pp\rightarrow ppK^{+}K^{-}$ reaction were
conducted at low excess energies by the collaborations ANKE~\cite{anke},
COSY-11~\cite{quentmeier,wolke,winter} and DISTO~\cite{disto}, and revealed a 
significant discrepancy between obtained excitation function
and theoretical expectations neglecting interaction of kaons in the final state.
The inclusion of the $pK^{-}$--FSI, based on factorization ansatz introduced by
the ANKE collaboration, reproduces the experimental data for 
excess energies down to Q~=~28~MeV, but underestimates the data very close
to threshold~\cite{anke,wilkin}. 
This indicates that in this energy region the influence of 
the $K^{+}K^{-}$ interaction may be significant.
Therefore, we have performed a more detailed analysis of the COSY-11 data
at excess energies of Q~=~10~MeV and 28~MeV including studies of the differential cross
section distributions~\cite{silar}.
Based on the mentioned factorization ansatz with additional term describing interaction in
the $K^+K^-$ system we compared the experimental event distributions
to the results of Monte Carlo simulations treating the $K^+K^-$ scattering length as an unknown parameter, which
has to be determined. Finally, based for the first time on the low energy $K^{+}K^{-}$ invariant
mass distributions and the generalized Dalitz plot analysis, we have estimated
the $K^+K^-$ scattering length~\cite{silar}.\\
Due to the low statistics the uncertainties of the determined scattering length are rather large,
therefore we decided to combine Goldhaber Plot analysis of COSY-11 data 
with fitting to the experimental excitation function near threshold.
We have constructed the $\chi^{2}$ statistic:
\begin{equation}
\chi^2\left(a_{K^+K^-},\alpha\right) = 2 \cdot \sum_i \, [\alpha N_i^s - N_i^e +  N_i^e \,
\text{ln}(\frac{N_i^e}{\alpha N_i^s})] + \sum_j\frac{\left(\sigma_{j}^{exp} - \alpha\sigma_{j}^{mod}\right)^2}{\left(\Delta\sigma_{j}^{exp}\right)^2}~,
\label{eqchi2_mh}
\end{equation}
where $N_i^e$ denotes the number of events in the $i$th bin of the experimental Goldhaber plot,
$N_i^s$ stands for the content of the same bin in the simulated distributions, $\sigma_{j}^{mod}$
denotes the $j$th calculated total cross section, $\sigma_{j}^{exp}$ is the $j$th experimental
total cross section measured with uncertainty $\Delta\sigma_{j}^{exp}$, and $\alpha$ stands for
the normalization factor.
\begin{figure}
\includegraphics[width=0.32\textwidth,angle=0]{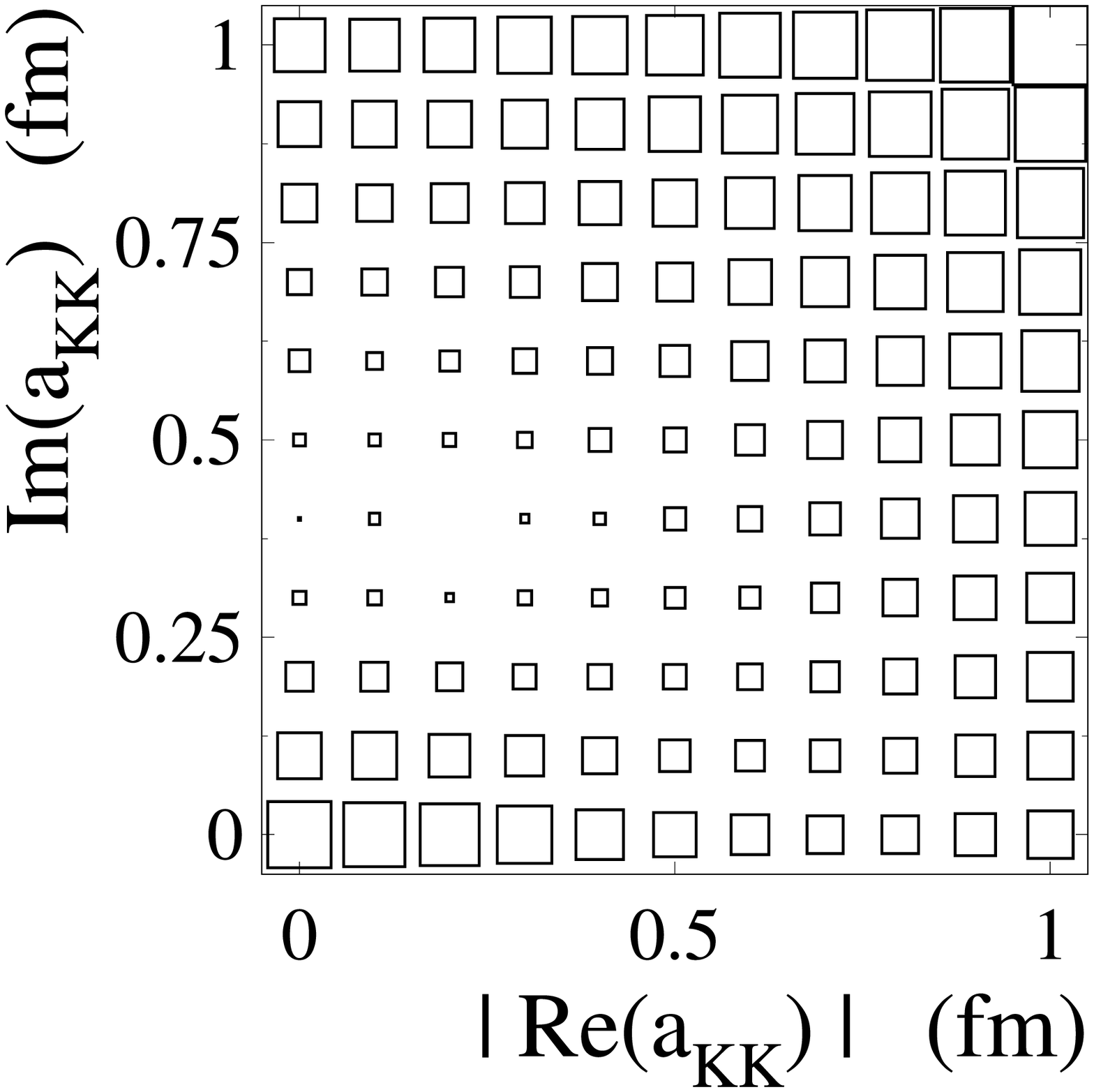}
\includegraphics[width=0.32\textwidth,angle=0]{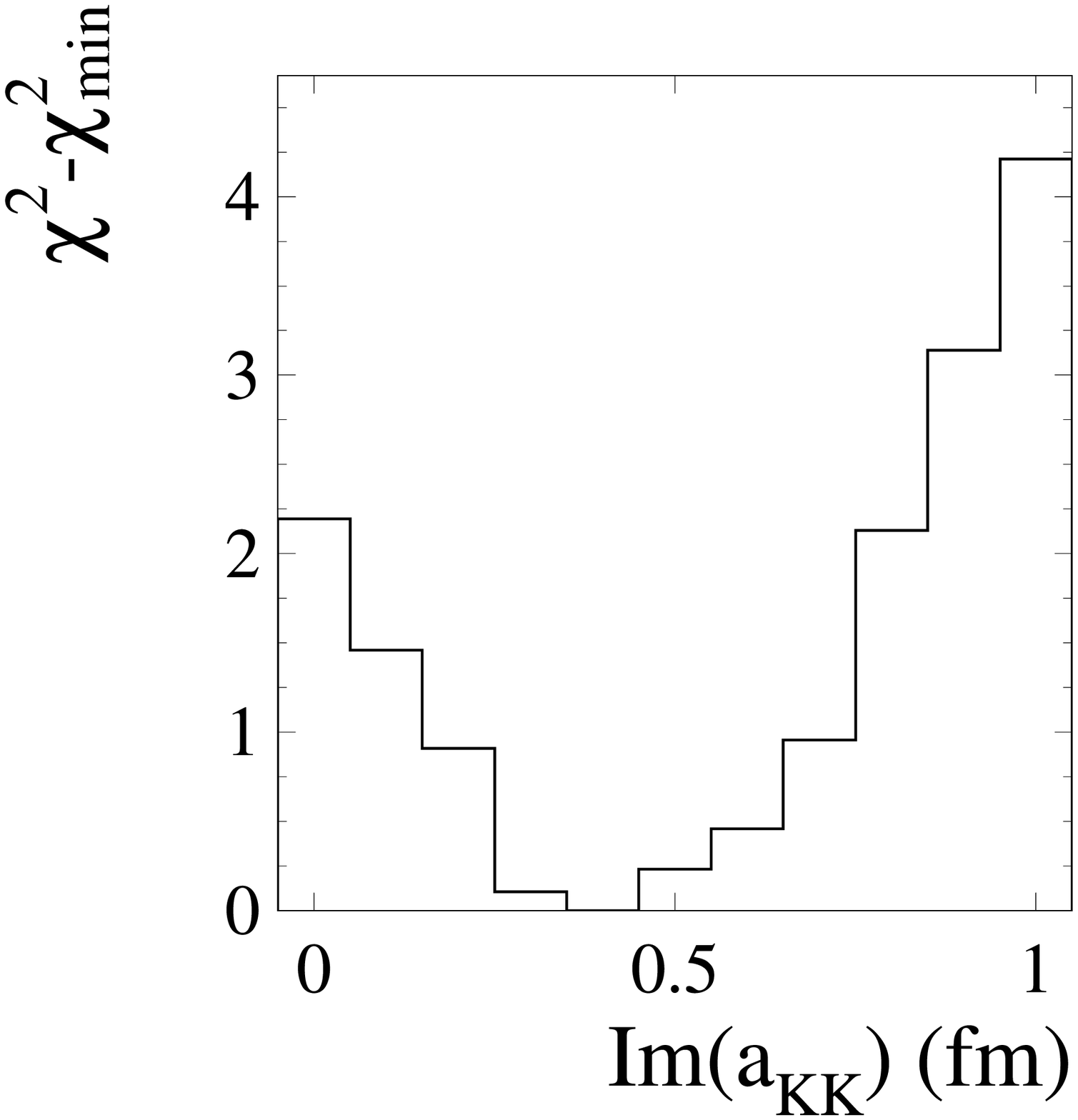}
\includegraphics[width=0.32\textwidth,angle=0]{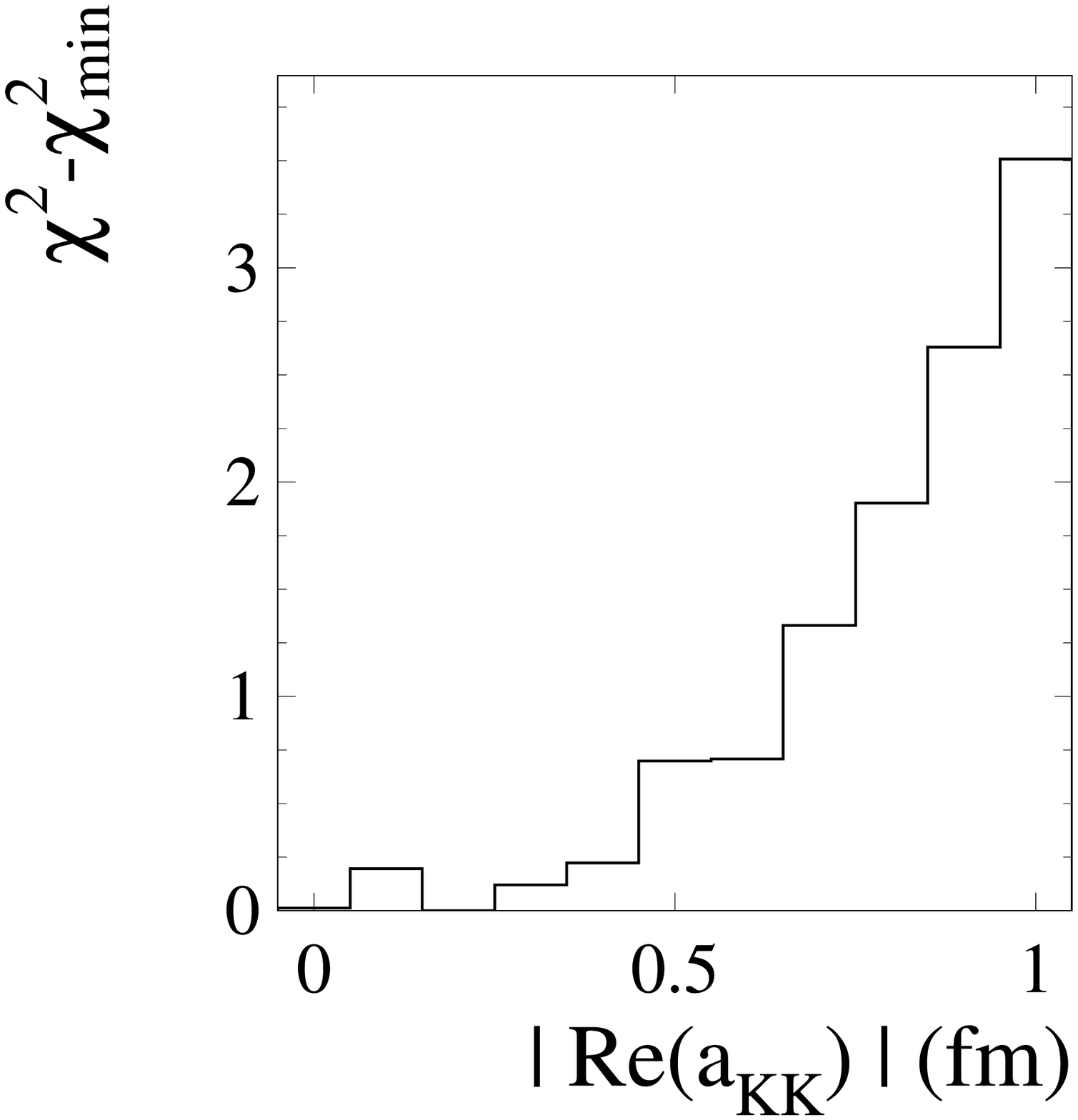}
\caption{
$\chi^{2}$~-~$\chi^{2}_{min}$ distribution as a function of $|Re(a_{K^{+}K^{-}})|$ and $Im(a_{K^{+}K^{-}})$.
$\chi^{2}_{min}$ denotes the absolute minimum with respect to parameters $\alpha$, $|Re(a_{K^{+}K^{-}})|$,
and $Im(a_{K^+K^-})$.}
 \label{chi2_1}
\end{figure}
The obtained preliminary $\chi^2$ distributions (suppressed by its minimum value) as a function of the real
and imaginary part of the $K^+K^-$ scattering length are presented in Fig.~\ref{chi2_1}.
The best fit to the experimental data corresponds
to $\left|Re(a_{K^{+}K^{-}})\right| = 0.2^{~+0.8}_{~-0.2}$~fm and $Im(a_{K^{+}K^{-}}) = 0.4~^{~+0.6}_{~-0.4}$~fm.
This preliminary result is by a factor of five more precise than the previous one based only on the analysis of the
differential cross section distributions~\cite{silar}.
%
\section{Conclusions}
The analysis of the $pp\rightarrow ppK^{+}K^{-}$ reaction measured by COSY-11 collaboration
at excess energy Q = 10 MeV and Q = 28 MeV  has been extended to the
determination of the differential cross sections in view of the $K^{+}K^{-}$ final state interaction.
To reduce the uncertainties on the $K^+K^-$ scattering length we have performed combined analysis of both
total and differential cross section distributions for the $pp\rightarrow ppK^{+}K^{-}$ reaction. The preliminary
result of the analysis is by a factor of five more precise, however the determined scattering length is still
consistent with zero within one standard deviation.

\section*{Acknowledgments}
The work was partially supported by the European Commission under the 7th Framework Programme
through the ''Research Infrastructures'' action of the ''Capacities'' Programme.
Call: FP7-INFRASTRUCTURES-2008-1, Grant Agreement N. 227431, by the PrimeNet,
by the Polish Ministry of Science and Higher Education through grant No. 1202/DFG/2007/03,
by the German Research Foundation (DFG), by the FFE grants from the Research Center J{\"u}lich,
and by the virtual institute ''Spin and strong QCD'' (VH-VP-231).


\begin{thebibliography}{00}    
\bibitem{Morgan}
D.~Morgan,~M.~R.~Pennington,~Phys. Rev.~D~\textbf{48}, 1185 (1993).
\bibitem{Lohse}
D.~Lohse~\textit{et al.},~Nucl. Phys. \textbf{A516}, 513 (1990).
\bibitem{Weinstein}
J.~D.~Weinstein,~N.~Isgur,~Phys. Rev.~D~\textbf{41}, 2236 (1990).
\bibitem{review}
P.~Moskal, M.~Wolke, A.~Khoukaz, W.~Oelert, Prog. Part. Nucl. Phys. \textbf{49}, 1 (2002).
\bibitem{cosy}
D. Prasuhn~\textit{et al.}, Nucl. Instrum. Methods Phys. Res.~A \textbf{441}, 167 (2000);
\bibitem{brauksiepe}
S.~Brauksiepe {\it et~al.}, Nucl. Instrum. Methods A \textbf{376}, 397 (1996).
\bibitem{quentmeier}
C.~Quentmeier~\textit{et al.},~Phys. Lett. \textbf{B515}, 276 (2001).
\bibitem{wolke}
M.~Wolke,~PhD thesis, IKP J{\"u}l-3532 (1997).
\bibitem{winter}
P.~Winter~\textit{et al.},~Phys. Lett. \textbf{B635}, 23 (2006).
\bibitem{anke}
Y.~Maeda~\textit{et al.},~Phys. Rev.~C \textbf{77}, 01524 (2008).
\bibitem{disto}
F.~Balestra~\textit{et al.},~Phys. Lett. \textbf{B468}, 7 (1999).
\bibitem{wilkin}C. Wilkin, Acta Phys. Polon. Supp. 2,  89 (2009); arXiv:0812.0098 
\bibitem{silar}
M.~Silarski~\textit{et al.},~Phys. Rev.~C \textbf{80}, 045202 (2009).
\end{thebibliography}
\end{document}